\documentclass{article}
\usepackage{spconf,amsmath,graphicx}
\usepackage{cite}
\usepackage{amsmath,amssymb,amsfonts}
\usepackage{algorithmic}
\usepackage{graphicx}
\usepackage{textcomp}
\usepackage{xcolor}
\usepackage{multirow}
\usepackage{comment}
\usepackage{subcaption}
\usepackage{stfloats} 
\usepackage{nopageno}
\usepackage{url}

\title{SEC2SEC CO-ATTENTION TRANSFORMER FOR VIDEO-BASED APPARENT AFFECTIVE PREDICTION}
\name{Mingwei Sun and Kunpeng Zhang}
\address{University of Maryland}
\begin{document}
%
\maketitle

\begin{abstract}
Video-based apparent affect detection plays a crucial role in video understanding, as it encompasses various elements such as vision, audio, audio-visual interactions, and spatiotemporal information, which are essential for accurate video predictions. However, existing approaches often focus on extracting only a subset of these elements, resulting in the limited predictive capacity of their models. To address this limitation, we propose a novel LSTM-based network augmented with a Transformer co-attention mechanism for predicting apparent affect in videos. We demonstrate that our proposed Sec2Sec Co-attention Transformer surpasses multiple state-of-the-art methods in predicting apparent affect on two widely used datasets: LIRIS-ACCEDE and First Impressions. Notably, our model offers interpretability, allowing us to examine the contributions of different time points to the overall prediction. The implementation is available at: \url{https://github.com/nestor-sun/sec2sec}.
\end{abstract}

\begin{keywords}
Co-attention, Transformer, Multimodal Learning, Affective Computing, Video Understanding
\end{keywords}
\vspace{-3mm}
\section{Introduction}
\noindent Video plays a crucial role in the field of human-computer interaction, and understanding human responses to video content is essential for designing and optimizing these systems. A key aspect of human interpretation is the perceived affect elicited by video content, as it can influence user engagement \cite{engagement1}, user trust in branded content \cite{trust}, and user behavior \cite{behavior} within these systems.

A video comprises two primary components: vision and audio. Each of these components has distinct influences on predictions. Furthermore, both visual and auditory elements can jointly impact predictions. Specifically, a good alignment between a visual component with its corresponding audio within the same time frame (e.g., within a second) creates a synergy that enhances viewers' perception of video. In the context of emotion perception, distinct combinations of audio and visual features may elicit different emotional states. For instance, a chilling image coupled with a slow audio tempo is likely to evoke a negative valence. Another crucial aspect that influences viewers' perception of a video is its sequential composition. Research indicates that people tend to recall the most recent information \cite{recency}, suggesting that later clips may carry higher weights when estimating affective responses.

In recent years, there has been significant attention given to video-based affective prediction. Studies have leveraged deep learning techniques to predict affect of video. For example, one study developed a convolutional neural network (CNN) to predict emotions using only images from videos \cite{cnn}. Previous research has emphasized the importance of various types of information in video-based affective prediction, including audio and visual features, temporal information, and audio-visual interactions. For instance, different attention mechanisms were explored, confirming the significance of capturing audio-visual interactions \cite{cross_modal_interaction}. However, despite these advances, affective prediction remains a challenging research topic, as existing methods have not yet fully captured all the relevant information present in videos.

In this paper, we tackle these challenges by developing a Transformer-based second-to-second (Sec2Sec) co-attention model to predict the video-perceived affective states. Specifically, we first implement a Transformer-based co-attention network extended from the work proposed by \cite{look_listen_attend} to understand the interactions between audio and vision. We further combine a Long Short-Term Memory (LSTM) module with such a co-attention network to capture the temporal information of videos at the second level. To do so, we first split each video into one-second video clips. We then feed each one-second clip into our designed co-attention network. The output of each video clip from the co-attention network is fed into an LSTM network sequentially. Lastly, we add a fully-connected feed forward (FC) for affective prediction. 

Our contributions are three-fold. First, we propose a novel Sec2Sec Co-attention Transformer for the affective prediction to capture all the necessary types of information. Second, we evaluate the performance of our proposed Sec2Sec Co-attention Transformer on two real-world datasets. Our extensive experimental results demonstrate that our approach outperforms several competitive baselines in affective prediction tasks. Additionally, we conduct interpretability analyses to assess the contributions of individual one-second video segments to the final predictions.

\section{Related Work}
\vspace{-3mm}

\noindent Our work is related to two streams of literature: audio-video representation learning and applications of Transformers. 

\begin{figure*}
        \centering
    \includegraphics[width=\textwidth]{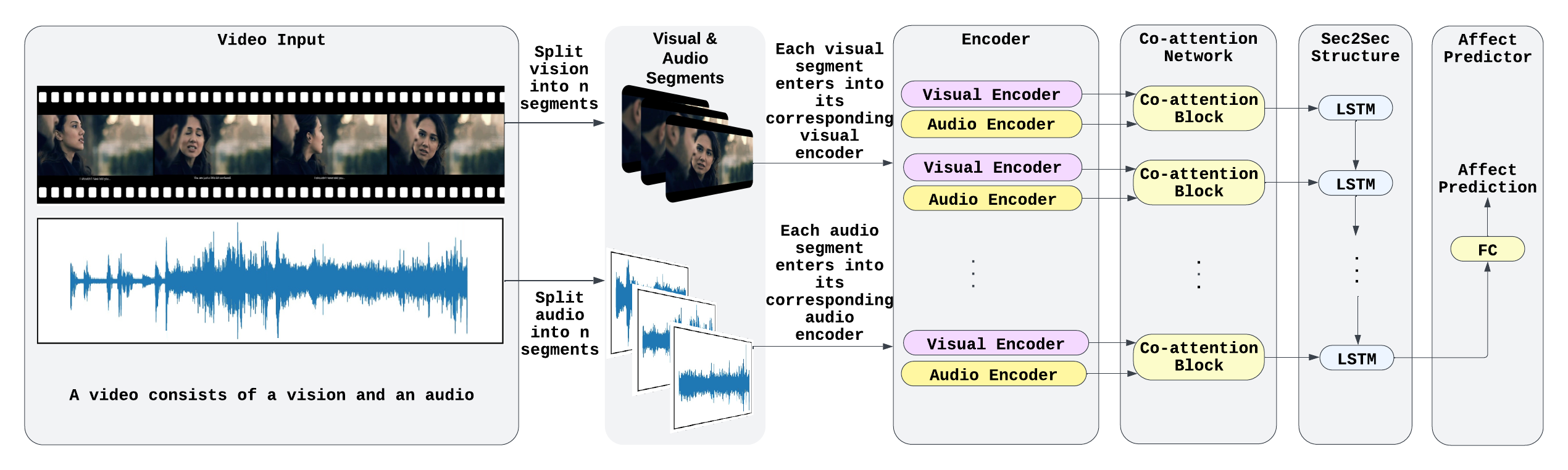}
    \vspace{-8mm}
    \caption{An overview of our proposed model: a Sec2Sec Co-attention Transformer.}
    \label{fig:architecture}
\vspace{-5mm}
\end{figure*}

\noindent \textbf{Audio-Visual Cross-Modal Learning}. The mainstream of audio-visual representation learning research is to predict the synchronization or correspondence of audio and visual streams in videos. Arandjelovic and Zisserman trained an audio-visual cross-modal network from scratch to predict video correspondence \cite{look_listen_learn}. Alwassel et al. used one clustered modality as a supervisory signal for another modality, and predicted correspondence between two modalities \cite{xdc}. Cheng et al. further developed three self-supervised co-attention-based networks to discriminate visual events related to audio events \cite{look_listen_attend}.  In addition, Kuhnke et al. proposed a two-stream aural-visual model to predict facial expressions in videos \cite{two_stream_aural_visual}.

\noindent\textbf{Transformers in Computer Vision}. Recently, Transformers have been increasingly applied to computer vision (CV) tasks as an alternative to CNN. ViT applies a Transformer model to linearly projected sequences of image patches to classify full images \cite{vit}. Swin Transformer improves ViT by introducing a hierarchical Transformer architecture and a shifted window scheme \cite{swin}. In order to classify video tasks, ViViT extends ViT by proposing two methods for embedding video samples: uniform frames sampling and tubelet embedding \cite{vivit}. Video Swin Transformer further extends Swin Transformer by introducing a 3D-shifted window-based multi-head self-attention module and a locality inductive bias to the self-attention module \cite{swin_video}. However, all these video-based analyses do not separate vision and audio and explicitly learn the joint effect on subsequent tasks, which is our focus in this study.

\vspace{-5mm}
\section{Method: Sec2Sec Co-attention Transformer}
\vspace{-3mm}

In this section, we introduce the proposed model. As depicted in Figure \ref{fig:architecture}, the proposed model consists of five steps: \textbf{Video Segmentation}: We first split each video into $n$ segments. \textbf{Encoder network}: The encoder network comprises a visual encoder and an audio encoder that extract visual and audio features using pre-trained ResNet networks. \cite{resnet}. \textbf{Co-attention block}: The co-attention block leverages Transformer \cite{attention_is_all_you_need} to model the interactions between visual and audio features. \textbf{Sec2Sec structure}: It captures the temporal information via an LSTM network. \textbf{Predictor}: The output from the LSTM network is fed into an FC network to make apparent predictions.

\noindent \textbf{Visual encoder}. To extract visual features, we first sample $m$ frames per segment. Each frame is represented by a color image with the Red-Green-Blue (RGB) channels. Like prior studies, we apply pre-processing to images, such as resizing, center cropping, and normalization. Thus, each visual part is represented in a 4-dimensional space (i.e., 3-dimensional RGB plus $m$ frames), which is fed into a pre-trained R(2+1)D ResNet model \cite{3dresnet}.

\noindent \textbf{Audio Encoder}. To extract audio features, we compute 2-dimensional Mel-Frequency Cepstral Coefficients (MFCCs) \cite{mfcc}, MFCC's first-order (delta coefficients), and second-order frame-to-frame time derivatives (delta-delta coefficients) from each audio clip. Therefore, the audio feature can be represented by combining three-channel MFCC features in which each channel is one type of coefficient. The three-channel MFCC features, treated as a special type of "image", are fed into a pre-trained ResNet \cite{resnet}.

\noindent \textbf{Co-attention block}. The extracted visual and audio features for each segment enter into two symmetric co-attention sub-blocks, visual and audio sub-blocks, to learn guided audio and visual representations. Each sub-block is built by combining a standard multi-head self-attention module with a multi-head co-attention module. A normalization layer (Norm), a residual connection and an FC network are applied.

In the visual sub-block, the extracted visual embedding from the visual encoder is first fed into a multi-head self-attention module to get the intermediate visual representation, $I_{v}$, embedding important visual information. Similarly, we can get the intermediate audio representation, $I_{a}$, in the audio sub-block. Specifically, $I_{v}$ and $I_{a}$ can be computed as follows:
\begin{equation} 
\begin{split}
I_{iv} = FC(Norm(MultiHead(z_{iv}, z_{iv}, z_{iv})) + z_{iv})\\
I_{ia} = FC(Norm(MultiHead(z_{ia}, z_{ia}, z_{ia})) + z_{ia}) \\
\end{split}
\end{equation}
where $z_{iv}$ and $z_{ia}$ denote the output features from the visual encoder and the audio encoder for segment $i$, respectively. 

Next, in the visual sub-block, $I_{ia}$  as key and value and $I_{iv}$ as query are passed into the multi-head co-attention module. In this way, we can enforce the visual sub-block to focus on the information related to audio. Similarly, in the audio sub-block, we feed $I_{iv}$ as key and value and $I_{ia}$ as query into the multi-head co-attention layer. Thus, the audio sub-block tends to focus on the information corresponding to vision. Hence, the final output features of vision and audio, $F_{iv}$ and $F_{ia}$, can be computed as:
\begin{equation} 
\begin{split}
F_{iv} = FC(Norm(MultiHead(I_{iv}, I_{ia}, I_{ia})) + I_{iv})\\
F_{ia} = FC(Norm(MultiHead(I_{ia}, I_{iv}, I_{iv})) + I_{ia}) \\
\end{split}
\end{equation}
Consequently, two sub-blocks focus on important information about themselves, as well as their relationships. Using such a mechanism, we capture the interaction between visual and audio components. Finally, we combine the guided visual and audio representation by applying an FC layer, computed as $F_{i} = FC(concat(F_{iv}, F_{ia}))$, which is the joint representation of vision and audio for each segment $i$. 

\noindent \textbf{Sec2Sec Structure}. To capture the temporal information in the video clip sequence, we feed the joint representation of each segment, $F_{i}$, from the co-attention block to an LSTM network. The LSTM network is defined as follows: 
\begin{equation}
\begin{gathered}
    u_{i} = \sigma(W_{Fu}F_{i} + W_{hu}h_{i-1}+b_{u})\\
    f_{i} = \sigma(W_{Ff}F_{i} + W_{hf}h_{i-1} + b_{f})\\
    o_{i} = \sigma(W_{Fo}F_{i} + W_{ho}h_{i-1} + b_{o}) \\
    \tilde{c}_{i} = tanh(W_{Fc}F_{i}+W_{hc}h_{i-1}+b_{c}) \\
    c_{i} = f_{i} \odot c_{i-1} + u_{i} \odot \tilde{c}_{i} \\
    h_{i} = o_{i} \odot tanh(c_{i})
\end{gathered}
\end{equation}
where $\sigma(\cdot)$ is an activation function. $\odot$ denotes the Hadamard product. $W$ and $b$ are weights and biases to be learned. $h_{i}$ denotes the hidden state at step $i$. $u_{i}$, $f_{i}$, $o_{i}$ and $c_{i}$ denote the update gate, forget gate, output gate and cell gate, respectively. 

\noindent \textbf{Predictor}. We apply an FC along with a sigmoid function to the output from the LSTM network at the last step to make affective predictions.

\vspace{-5mm}

\section{Experiments}
\vspace{-3mm}

\subsection{Dataset and Evaluation}
\noindent To evaluate the effectiveness of our proposed model, we utilize two publicly available datasets: LIRIS-ACCEDE \cite{liris} and First Impressions \cite{first_impressions}. LIRIS-ACCEDE contains $9,800$ videos extracted from $160$ films. In this paper, we adopt a binary classification approach based on the existing literature \cite{binary_label}. For evaluation, we adopt two standard metrics for emotion classification tasks (valence and arousal), accuracy and $F1$ score. First Impressions is widely utilized in the field of apparent personality analysis consisting of 10,000 labeled clips extracted from over 3,000 YouTube videos. Since the personality traits are continuous between 0 and 1, we use the mean accuracy as the evaluation metric, computed as, $MeanAccuracy_{t} = \frac{1}{N}\sum_{i=1}^{N}(1-|y_{it} - \hat{y}_{it}|)$, where $y_{it}$ is the ground truth value for the $i$th video sample and $t$th personality trait, and $\hat{y}_{it}$ is the predicted value for the same video sample and trait. $N$ is the total number of predicted videos.

\subsection{Implementation Details}
\noindent We train all models on an NVIDIA GeForce 3090 24GB GPU with 250 epochs. We set up an early stopping mechanism, where the training stops if the validation loss increases for 5 consecutive epochs. We use the grid search strategy to find a relatively optimal set of hyperparameters. In each experiment, we use the model with the best validation accuracy to report results on the holdout testing set. 

\subsection{Baselines}
\noindent We evaluate the performance of our proposed model (called \textbf{Sec2Sec SA-CA}) with several state-of-the-art methods. \textbf{CMA}, \cite{look_listen_attend}, \textbf{AVM} \cite{two_stream_aural_visual}, \textbf{ViT} \cite{vit}, \textbf{ViViT} \cite{vivit}, \textbf{ViT-ViViT} \cite{vit, vivit}: We implement a bi-modal (audio and vision) network by combining ViT and ViViT to extract audio and visual features, respectively. We also add a co-attention network as another baseline and 3 variants of our model for comparison to understand the role of each design in our model (e.g., uni vs. bi-modal, co-attention): \textbf{Co-attention} (bi-modal), \textbf{Sec2Sec Vision} (uni-modal), \textbf{Sec2Sec Audio} (uni-modal) and \textbf{Sec2Sec SA-SA} (bi-modal with self-attention).

\begin{table*}[t]
\centering
\begin{center}
\caption{Performance comparison of our model with baselines for arousal and valence prediction.}
\resizebox{\linewidth}{!}{
\label{table:arousal_result}
\begin{tabular}{c |c  | c | c  c  c | c c c}
\hline 
&        &          &   \multicolumn{3}{c}{Arousal}  &   \multicolumn{3}{|c}{Valence}  \\
& Method & Input & Accuracy & $F1$ Score & Avg Training Time & Accuracy & $F1$ Score & Avg Training Time \\ 
&        &          &      &  &  Per Epoch (min) &    &   &  Per Epoch (min) \\
\hline
\multirow{6}{*}{Baselines}
&ViT\cite{vit} & Audio & 0.7823 & 0.8768&1:55 & 0.7022 & 0.8154&1:52\\
& ViViT \cite{vivit} & Vision & 0.7853 & 0.8795 & 1:32 & 0.7002 & 0.8234 & 1:29\\
& CMA \cite{look_listen_attend} & Audio and Vision & 0.5680  & 0.6768 & 4:50& 0.6078  & 0.7033 & 6:05  \\ 
& AVM \cite{two_stream_aural_visual} & Audio and Vision & 0.7756 & 0.8722 & 4:49 & 0.7205  & 0.8287 & 4:49\\

& ViT-ViViT \cite{vit, vivit} & Audio and Vision & 0.7517 & 0.8541 & 3:31&  0.6901 & 0.8009& 3:32\\

 & Co-attention & Audio and Vision & 0.5599  & 0.6603 & 4:50 &  0.5864  & 0.6688 & 4:50\\ \hline
\multirow{4}{*}{Variants of Ours}  & Sec2Sec Audio & Audio & 0.7832	 & 0.8780 & 1:51 & 0.7021  & 0.8191 & 1:50 \\ 
& Sec2Sec Vision & Vision &0.7766  & 0.8733 & 1:20& 0.6970  & 0.8179 & 1:20  \\
& Sec2Sec SA-SA &Audio and Vision & \textbf{0.7990}  & \textbf{0.8876} & 2:17 &  0.7047 & 0.8179 & 2:14 \\ 
&Sec2Sec SA-CA &Audio and Vision  &0.7949 & 0.8840 & 2:14&  \textbf{0.7322} & \textbf{0.8372}& 2:15 \\
\hline
\end{tabular}
}
\end{center}
\vspace{-0.5cm}
\end{table*}

\begin{table*}[ht]
\centering
\begin{center}
\caption{Performance comparison of our model with baselines for personality prediction. }

\label{table:personality_result}
\resizebox{\linewidth}{!}{
\begin{tabular}{c | c | c | c c c c c c }
\hline 
& Method & Modality & Agreeableness & Conscientiousness & Extraversion & Neuroticism & Openness & Avg Training Time  \\ 
&        &          &          &    &   &   &  & Per Epoch (min) \\
\hline
\multirow{6}{*}{Baselines} 
 &ViT\cite{vit}  & Audio &0.891	& 0.879	& 0.885	& 0.886 & 0.887 & 1:03\\
& ViViT\cite{vivit} & Vision & 0.894 &	0.876 &	0.878 &	0.877	& 0.883 & 4:48\\
&  CMA \cite{look_listen_attend} & Audio and Vision& 0.894 &0.882 &	0.887 &	0.889 &	0.890 & 2:24  \\ 
 & AVM \cite{two_stream_aural_visual} & Audio and Vision&  0.894	& 0.881 &	0.889 &	0.887 &	0.893 & 2:17 \\ 

& ViT-ViViT \cite{vit, vivit}& Audio and Vision&  0.897 & 0.882 &	0.886 &	0.889 &	0.892
 & 6:15\\
& Co-attention &  Audio and Vision& 0.890 &	0.883 &	0.883 &	0.888 &	0.892
& 1:29 \\ \hline
\multirow{4}{*}{Variants of Ours}  &  Sec2Sec Audio &Audio & 0.895 & 0.878 &0.884	&0.881 &	0.888 & 0:23 \\ 
& Sec2Sec Vision &Vision& 0.893	& 0.876 &	0.879 &	0.877 &	0.883 & 0:23 \\

& Sec2Sec SA-SA & Audio and Vision&  0.895 & 0.878 & 0.884 & 0.881 & 0.888 &  0:49 \\

& Sec2Sec SA-CA & Audio and Vision&  \textbf{0.898} & \textbf{0.891} &	\textbf{0.892} &	\textbf{0.892} &	\textbf{0.896}
& 1:27 \\ \hline 
\end{tabular}
}
\end{center}
\vspace{-7mm}
\end{table*}

\subsection{Results}
\noindent \textbf{Overall performance}. Table \ref{table:arousal_result} presents the experimental results for arousal and valence. Our proposed Sec2Sec models achieve the best performance on two evaluation metrics for arousal prediction. They surpass three bi-modal baselines (audio and vision) methods and the co-attention approach in terms of accuracy, F1 score and efficiency, demonstrating the benefit of incorporating LSTM (Sec2Sec) into the video understanding framework. It also outperforms Sec2Sec Audio and Sec2Sec Vision, indicating that using both audio and visual components is more effective than using either modality alone. Moreover, Sec2Sec SA-SA and Sec2Sec SA-CA obtain comparable results, suggesting that the interactions between audio and visual features is not essential for predicting arousal. It is noteworthy that ViT-ViViT performs worse than ViT and ViViT, indicating that a single FC layer fails to adequately capture the interaction between audio embeddings and visual embeddings. We have similar observations for valence prediction, in terms of performance comparison with baselines. However, Sec2Sec SA-CA outperforms Sec2Sec SA-SA, indicating the usefulness of the co-attention mechanism at predicting valence. 

Table \ref{table:personality_result} presents the experimental results for personality predictions. The proposed Sec2Sec SA-CA method consistently outperforms three bi-modal baseline methods and the co-attention method in all five personality label predictions. This confirms the effectiveness of the Sec2Sec structure and highlights the importance of spatio-temporal information. Furthermore, CMA consistently outperforms other baselines, demonstrating the importance of visual and audio information as well as audio-visual interactions captured by cross-modal attention. Notably, Sec2Sec SA-CA requires less training time than the baselines, illustrating the efficiency of the Sec2Sec structure. Additionally, Sec2Sec SA-CA outperforms both Sec2Sec Audio and Sec2Sec Vision, indicating that bi-modal information is more effective than using either modality alone for personality predictions. Finally, the superior performance of Sec2Sec SA-CA over Sec2Sec SA-SA demonstrates the ability of the co-attention mechanism to capture rich interaction information between audio and vision.

\begin{figure}[t]
    \centering
    \begin{subfigure}[b]{0.21\textwidth}
         \includegraphics[width=\textwidth]{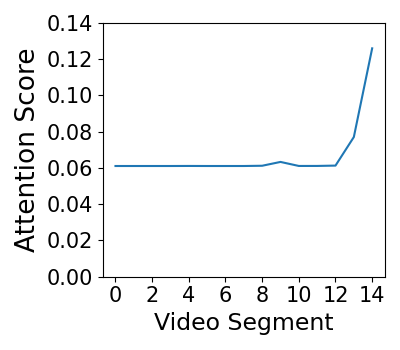}
         \caption{Valence}
         \label{fig:lstm_valence}
    \end{subfigure}
    \hfill
    \begin{subfigure}[b]{0.21\textwidth}
         \includegraphics[width=\textwidth]{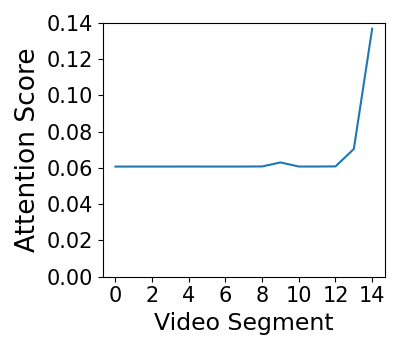}
         \caption{Arousal}
         \label{fig:lstm_arousal}
    \end{subfigure}
    \caption{The LSTM Attentions for LIRIS-ACCEDE.}
    \label{fig:liris_lstm_attention}
    \vspace{-8mm}
\end{figure}
\noindent\textbf{Model Interpretability}. We assess the contribution of each video segment (i.e., every one-second clip) to emotion prediction by substituting LSTM with an attention-based LSTM proposed by \cite{lstm_attention}. After training, we obtain the learned LSTM attention values. The attention values of each video segment for valence and arousal are plotted in Figure \ref{fig:lstm_valence} and \ref{fig:lstm_arousal}, respectively. Similar patterns are observed in both figures. They showcase that the emotion prediction power reaches the highest for the last 3 seconds of the video, suggesting that emotions are mostly influenced by the last 3 seconds. Moreover, the impact of video segments increases as they approach the end of a video.
\vspace{-5mm}

\section{Conclusion}
\vspace{-3mm}

\noindent This study introduces an innovative Sec2Sec Co-attention Transformer model for perceived affect prediction in videos. Our approach harnesses the power of pre-trained ResNet networks, LSTM, and a co-attention mechanism to effectively encode and integrate multimodal features. The experimental results underscore the efficiency of our Sec2Sec structure and the significance of inter-modal interaction in affective prediction. In addition, we present an attention-driven LSTM technique to examine the impact of each second clip within a video on the overall affective prediction. 

\vspace{-5mm}
\bibliographystyle{IEEEbib}
\bibliography{ref.bib}

\end{document}